\documentclass[a4paper]{jpconf}
\usepackage{graphicx}
\begin{document}
\title{Qualitative insights on fundamental mechanics}

\author{Ghenadie N. Mardari}

\address{Rutgers University, Piscataway, NJ 08854, U.S.A.}

\ead{g.mardari@rutgers.edu}

\begin{abstract}
The gap between classical mechanics and quantum mechanics has an
important interpretive implication: the Universe must have an
irreducible fundamental level, which determines the properties of
matter at higher levels of organization. We show that the main
parameters of any fundamental model must be theory-independent. They
cannot be predicted, because they cannot have internal causes.
However, it is possible to describe them in the language of
classical mechanics. We invoke philosophical reasons in favor of a
specific model, which treats particles as sources of real waves.
Experimental considerations for gravitational, electromagnetic, and
quantum phenomena are outlined.
\end{abstract}

\section{Introduction}
Science is built on the assumption of determinism. It makes no sense
to even begin to explain a phenomenon, unless we believe that it is
governed by some sort of law. And if we believe that laws are truly
at work (as proven by our successes in science), how can we not also
believe that we can describe them in full? Indeed, the temptation is
high, and many philosophers have tried to come up with the ultimate
description of the Universe. However, it is very difficult to find
necessary links between general models and their particular
materialization. A model that is sufficiently abstract to explain
all the laws that are conceivable can lead to many possible worlds.
Why do we have one manifestation and not the other? On the other
hand, any particular state could have emerged from various initial
configurations. Which is the true one? Philosophers have tried to
overcome these hurdles without lasting success, and now, it seems,
it is the turn of mathematicians. The ability of mathematical models
to yield predictions about real phenomena has mesmerized many great
scientists. It has fostered a tacit belief in some sort of mystical
relationship between math and physical reality. And yet, the path to
final answers has not become much easier. Mathematical models are by
definition abstract. They have many types of consequences, but they
are impartial to all of them. Hence, quantitative models have also
gotten stuck in a multiplicity of approaches, without any obvious
standard for discrimination among them. It seems fair to conclude
that the ultimate solution, if it is at all possible, will not come
by way of pure abstraction, mathematical or otherwise. Some kind of
independent standard must be introduced, in order to bypass
irrelevant levels of abstraction. Empirical observations are helpful
for this task, but they require tremendous efforts. If qualitative
analysis could offer plausible shortcuts, by showing were to look
for evidence in the first place, the project might be simplified
dramatically. In this presentation, we shall discuss a possible way
to identify the basic principles of material organization with
qualitative methods.

Every physical observation limits our theoretical choices in one way
or another. In plain language, the fact that the Earth has mountains
constrains our options to the class of models that can predict the
development of planets with mountains on them. However, this
limitation is not sufficient for our purposes. We are looking for
some kind of process, whose properties are complex enough to require
a very specific type of explanation. Hence, there are many ways to
describe the possible emergence of matter, just like there are many
possible ways to describe the emergence of consciousness. It is only
the limits of our imagination that can slow the flow of hypotheses
on these issues. However, the exact interplay between consciousness
and matter is not a trivial matter. It involves so many
difficulties, that only a few scenarios can be plausible. In short,
the emergence of a material Universe, in which self-conscious
subjects can evolve consistently in well-defined contexts, can only
happen in a limited number of ways. If it were possible to identify
the fundamental properties, which allow for the possible existence
of rational beings, a lot of irrelevant models could be discarded.
This argument may sound similar to the anthropic principle, but our
angle is slightly different. Instead of trying to explain the
reasons for the actual manifestation of the Universe, we aim to
identify the fundamental unobservable processes that must be
inseparable from it.

A useful technique in this regard is to focus on the problem of
structural causality. Is the macroscopic manifestation of the
Universe determined by its microscopic properties? There are
sufficient scientific observations to warrant a positive answer. If
so, how many levels of organization are there? Does the Universe
have infinitely many structural levels, or is there a fundamental
level where all reductionism must stop? In order to answer this
question, we must ask another. What would the Universe look like, in
terms of its causal mechanisms, if it had infinitely many levels of
organization? From our point of view, the answer is straightforward:
an infinite number of levels translate into infinite reductionism,
without any break in causality. Every level of organization must be
neatly predicted by the properties of the lower level, which is
predicted by a still lower level, and so on without end. On the
other hand, what if the Universe had a fundamental level, which
determined the properties of all superior levels, without being
itself reducible to an inferior level? The implication is again
unavoidable: a fundamental level must determine the manifestation of
all laws, without being itself determined by them. Therefore, its
processes should obey a special set of laws. In other words, there
should be an unavoidable gap between the mechanics of fundamental
processes and that of all other levels of material organization.
When we compare the known properties of our Universe with the two
descriptions outlined above, it becomes apparent that only one of
them is relevant. Quantum mechanics is defined by its odd set of
properties, which do not obey the rules of classical processes. The
unavoidable gap between classical mechanics and quantum mechanics is
a crucial indicator. It shows that our Universe must have a
fundamental level, which determines the laws of all types of
material interaction, and which cannot be reduced to any lower level
of organization. Moreover, the properties of this fundamental level
must be such as to make the interaction between mind and matter
possible, as suggested in the previous paragraph.

What are the general properties of a fundamental level? By
definition, it cannot be reducible to lower levels, and yet it must
have some kind of properties. If it had none, it would not be able
to determine the properties of superior levels of material
organization. Moreover, it must determine those properties, without
being in turn determined by them. If not, causality would fail.
Consequently, the properties of the fundamental level cannot be
determined from inside the Universe. This means that they should
also be theory-independent in our models. Accordingly, we do not
need to try to explain the total amount of matter in the Universe,
or the exact value of Planck's constant (why this value, and not
some other?). Neither do we have to explain why the speed of light
is constant, or why it has its exact value. Properties like this
cannot be determined from internal processes in a Universe with a
fundamental level. In order to understand our world properly, we
just need to start from the fact of their existence and look for an
underlying mechanism that could incorporate them. The only
constraint is that this mechanism must determine the nature of all
other processes.

It may seem that we have excessive freedom in choosing the likely
properties of the fundamental level. However, if it is to lead to
self-conscious beings endowed with freedom, the options are quite
limited. In a separate paper (currently prepared for publication in
a philosophical journal) we have shown that the following set of
properties fit the task in a verifiable manner. Firstly, the
Universe must have a well-defined and constant amount of matter.
Secondly, it must contain a finite number of discrete active
entities (particles), which must also be identical. These entities
must move in discrete steps, producing physical effects at every
step. Thirdly, this motion and its effects must occur in some sort
of elastic medium, via mechanisms that are intelligible in terms of
macroscopic analogies. Without claiming that only these properties
can lead to a Universe like ours, it is our conclusion that they are
essential for the emergence and development of free self-conscious
beings in material environments. Any substantial modification to
their general description must be justified by showing that it does
not undermine this function. In the strongest formulation of our
argument, the mere existence of human beings requires the existence
of such properties. If the evidence were to contradict their
existence, the implication would be that the experienced world is
illusory (or else, a flaw in our analysis of self-consciousness
would have to be exposed).

The general properties described above can be used to develop a
detailed model, in order to verify their implications. Given the
constraint to have a fundamental level with theory-independent
properties, it is most convenient to use the analogy of a virtual
environment, in which the elementary entities come pre-programmed
with specific qualities. This approach belongs to the class of
models with multiple universes, in which the fundamental properties
of ``offspring'' self-enclosed realms are determined by the effects
of ``parent'' environments. Accordingly, the causes of elementary
processes cannot be properly described from inside. Yet, these
processes are expected to determine the nature of all the other
observables in the Universe. Hence, particles are described as
moving through the medium at constant rates, producing discrete
excitations (plucks) at every step. The unit of action,
corresponding to these constant energy inputs, must be equal to
Planck's constant. These excitations are described as the only
source of energy in the Universe, and the only way for the particles
to have effects on each other. For consistency, an additional rule
is required to rule out direct collisions between them. In the
absence of external influences from other sources, the trajectory of
any elementary particle must be rectilinear. The speed of
displacement, in this case, would have to correspond to the speed of
light. However, note that the medium is elastic and there is no
fundamental rigid unit of distance. The primary constant is the
number of steps in fixed units of absolute time. In other words,
energy is the fundamental constant in the model, as it is being
constantly produced in identical increments by all particles. These
identical particles are supposed to enter into various associations
on the basis of their mutual effects, producing all known subatomic,
atomic, molecular components and so on. Again, the symmetries of the
Universe are not expected to evolve from a hypothetical point of
conception in the past, but follow directly from the properties of
its fundamental constituents.

The model described above might be easier understood as a literal
representation of the Huygens Principle. As a reminder, Huygens
showed that every point on the front of a wave can be described as a
source of waves. This approach had its difficulties when applied to
electromagnetic phenomena, because it could not explain the
directionality of waves, or their quantized nature. In contrast,
here we assume the operation of elementary entities, which have
well-defined directions of motion and produce the waves in discrete
identical steps. According to this description, all fields can be
interpreted as wave-patterns in the medium, produced by the
longitudinal and transverse waves of the sources. The pure form of
these waves would be detectable as the zero-point field. They could
further assume the shape of electromagnetic fields, when emitted by
longitudinal particle associations. Static fields (electric or
magnetic) would be produced by transverse associations of elementary
particles. Finally, the gravitational force would have to be a
complex manifestation of electric properties. In plain terms, the
photons are supposed to look like trains of elementary particles,
where frequency is determined by the average interval between
elementary units. Subatomic particles would resemble micro-galaxies
of elementary constituents. It is of note that two symmetric
associations, which orbit each other, may cancel each other's field
in their own frame of reference, but not in all other frames that
are in relative motion. This residue of electric force would have to
manifest as gravity via dynamic effects, according to our
interpretation.

This description, though unorthodox, is not without precedent in the
literature. The hypothesis of sub-quantum mechanics, based on
classical interactions between identical constituents, was formally
investigated by G. Kaniadakis [1]. The wave aspect of our model
resonates with Theocharis' hypothesis about Maxwell fluxes [2],
though his qualitative considerations are quite different. Some
aspects of this approach might even resemble A.O. Barut's
self-energy model [3]. Common points with other models no doubt
could be found. Hence, it should be possible to work out the formal
details of our model in order to prove consistency with verified
elements of other theories. However, this report is only about the
efforts to overcome the early stages of theory-building. Before
proceeding to quantitative analysis, we need to be sure that the
qualitative implications are valid enough to make it worthwhile.
Consequently, we shall discuss the main features of this model, the
most likely objections to them, as well as the most important
experimental considerations. We shall argue that our description
makes several new predictions, which are highly consistent with the
empirical record. In some cases, verifiable implications will be
followed by a call for feed-back from the audience, especially when
the evidence is interesting but sketchy.

\section{Overcoming possible objections}

The hypothetical existence of a fundamental level with
theory-independent properties is not guaranteed to have scientific
value. It is worthy of attention only if it uncovers new phenomena
in the Universe, without contradicting existing knowledge. These
qualities are not immediately obvious, considering that energy
conservation does not seem to be obeyed in the preceding
description. Moreover, the postulation of a real medium appears to
be in conflict with well-known tests of invariance, in particular
with the Michelson-Morley experiments. Therefore, it is important to
address these concerns before discussing other aspects of this
model.

The principle of energy conservation is not above scientific
explanation. Ideally, its operation should be reducible to some sort
of causal relationship. The question is: should the causal mechanism
itself unfold in a manner that displays energy conservation? If it
did, would it really be the determining cause? In our opinion, the
cause should determine the effects without being determined in turn
by them. Circularity is highly undesirable in any good theory. This
position appears to be shared by other approaches as well. Hence,
the Big Bang is supposed to have produced all the matter in the
Universe from a theory-independent state. All of the known laws are
supposed to break down in the proximity of that original point.
There can no verifiable cause for the Big Bang, and none is being
promoted by scientists. In short, the source of energy in the
Universe is not explained in any relevant way. Similarly, the
hypothesis of a fundamental level of matter cannot propose a theory
for the source of energy. Instead of suggesting that it came at once
from a single point, this approach merely requires a process of
constant generation. Thus, the problem is not to prove that such a
mechanism is possible. The relevant part is to show that it can lead
to the manifestation of energy conservation, as it is observed on
all other levels of material organization.

As it turns out, energy conservation follows naturally from the
described model. The specified medium is supposed to be fundamental,
in the sense that all matter is embedded in it. Therefore, it cannot
have any kind of rigid barriers to serve as wave breakers. All the
waves must propagate away from their floating sources, without ever
being reflected, or refracted. Assuming that the waves are produced
at constant rates by the sources, and that they spread outwards at
constant rates in a homogenous medium, it follows that any finite
volume around a source must contain a constant amount of wave
energy. Note that wave energy is supposed to be the only kind of
energy in the Universe. Therefore, constant generation of waves can
only produce the appearance of energy conservation. In any finite
volume of space, which contains a finite number of sources, the
total energy will have minor fluctuations around a constant finite
value, in direct proportion to the number of sources. Energy
conservation is thus interpreted as the conservation of sources. All
transformation of energy must be reducible in one way or another to
a change in the pattern of interaction among sources. This
description makes it immediately obvious why the energy of a very
long photon is proportionate to its frequency multiplied by Planck's
constant. Frequency is determined by the number of elementary
sources in any segment, which produce waves in constant discrete
excitations, as they pass by a target or detector. Similarly,
subatomic particles have energies that are also proportional to the
number of elementary constituents, which explains their De Broglie
wavelength. Though, in these cases the excitations are produced via
circular patterns of motion, around a common center. Elementary
sources must always make a constant number of plucks on the medium.
In complex associations their mutual effects induce zitterbewegung,
which results in subluminal speeds of displacement relative to the
medium. This description can also explain the emergence of other
phenomena, such as inertia and variable kinetic energy.

Now, what about the Michelson-Morley experiments? Didn't they put to
rest all the models that involve classical mediation at the
fundamental level? Despite the widespread support for such a
position, it is not exactly accurate. Michelson and Morley tested
the assumption that light is a wave, propagating inside a physical
medium [4]. It is only by virtue of such an assumption that the
frequency of light should be Doppler shifted, when the source of
light is in motion relative to the medium. Yet, this assumption does
not apply to the model introduced above. If light consists of
longitudinal associations of elementary particulate sources, which
emit waves at every step, then the speed of light is the speed of
particles. The latter depends on the energy density of the medium,
which explains why gravity could act as a refraction coefficient for
light, inducing local invariance. According to this interpretation,
Michelson and Morley tested the speed of sources. The speed of waves
has yet to be determined, because it reflects an independent
process. What is the phenomenon that corresponds to these
fundamental waves? It has to be the fields that surround each
source. Therefore, in order to determine the speed of these waves,
and test them for invariance, the experiment should be aimed at
detecting the speed of propagation of changes within static fields
(electric, magnetic, or gravitational). Our approach survives the
objection of invariance by making a very interesting novel
prediction, which is also easily verifiable.

After considering several alternatives, we came to the conclusion
that the elementary sources must make two types of waves:
longitudinal and transverse. Longitudinal (electric) waves must
propagate in the direction of motion the sources, in opposite
directions, whereas transverse (magnetic) waves would have to be
orthogonal to them. In contrast to the traditional interpretation,
which describes electromagnetic waves in terms of mutually
generating fields, we just assume that sources are always generating
their field as described. Relative to a stationary detector, the
passage of a train of sources must produce the observation of an
oscillating field. If so, the waves that constitute the
electrostatic and magnetic fields should not be Lorenz invariant. An
important problem is to identify their speed. It has to be finite,
but there is no reason to expect it to be equal to the speed of
light. In fact, we believe that it should be superluminal, or else
it is very difficult to explain quantum interference. Very short
pulses of light, propagating simultaneously, would be unable to
interfere with each other if their waves had the speed of light,
because the waves would never be able to catch up with the adjacent
pulses. Moreover, in inertial systems moving at high speeds relative
to the medium, interference would have to be highly anisotropic,
contrary to known observations. A similar argument can be invoked to
show that gravitational redshift would be impossible within the
terms of this model, unless the speed of gravitational waves (which
is the speed of electrostatic waves) was higher than the speed of
light. Again, the waves would never be able to catch up with the
photons, in order to affect their frequency. (In this context, the
field is supposed to consist of many waves, constantly propagating
away from the source).

An obvious objection to the previous conclusion is that superluminal
waves would have been detected by now, had they been so common.
However, the usual interpretation of static fields was not conducive
to such discoveries, in our assessment. Many scientists believe that
there is nothing happening inside a static field. How can anything
propagate if it is static? Others believe that static effects are
instantaneous all over the universe. Yet another group of physicists
appear to believe that such effects must obey the speed of light.
The most common interpretation is that a changing electric or
magnetic field must instantly generate an electromagnetic wave,
which means that changes would propagate at the speed of light.
However, even if the radiation does propagate at the speed of light
(which we do not dispute), the question that we must answer is the
speed of changes inside the field, while it generates the
electromagnetic effect. It may be relevant to note here that
electromagnetic radiation is always due to the release of elementary
sources in our model, which explains why it has the speed of light.
The sudden change of a field due to a discharge, or due to a
magnetic spike is an independent effect, which must be investigated
exhaustively. If such a phenomenon were to be verified, the
mentioned hypotheses concerning fundamental mechanics would be
confirmed beyond reasonable doubt.

It seems that magnetic fields are the best candidates for initial
tests, because they are easier to manipulate. A very strong magnetic
pulse could be produced simply by discharging a capacitor through an
electromagnet. In order to test the speed of the pulse, an
experimenter would have to place two detectors in the same
direction, at different distances from the source. Knowing the time
of detection and the distance between detectors, it should be easy
to calculate the speed of the pulse. The problem, of course, is to
achieve the high accuracy that would be required for this task.
Rather than using clocks to register the time of detection directly,
it may be more convenient to have some electric switch connected to
the detector. This way, the effect of the discharge on the detector
could be accompanied by the release of a pulse of light. The pulses
of light from both detectors should be aimed towards a
photodetector, placed in the direction of propagation of the
magnetic pulse. Hence, the magnetic pulse would reach the first
detector, triggering a co-propagating pulse of light. Then it should
reach the second detector, triggering the second pulse of light. If
the speed of light is equal to the speed of the magnetic pulse, then
the two pulses of light should get to their detector simultaneously.
If the speed of the magnetic pulse is superluminal, the second pulse
of light should be the first to arrive at the detector. The distance
between magnetic detectors should be large enough to overcome any
uncertainty in the time of emission of the pulses of light. The
interval between photonic detections can be measured with a
multi-channel analyzer, or with interferometric methods. Once the
speed of the magnetic pulses is verified conclusively, different
experiments can be devised to test it for invariance.

In addition to this chief prediction about the speed of waves, the
model presented above has several other implications concerning
gravitational, electromagnetic, and quantum effects. In many cases,
experimental evidence is already available. Where this is the case,
it seems that the model is compatible with the data. The most
interesting findings known to us are outlined in the following
sections of this presentation.

\section{Relevant gravitational effects}

The assumption that fundamental entities can only produce a fixed
number of excitations per unit of time implies that all of them can
serve as clocks for each other. This provides a model for explaining
relativistic effects as well. The epicyclical motion of elementary
entities within subatomic particles depends on the speed of motion
of the association through the medium. At high velocities, more
units of action are required for the completion of each cycle and a
smaller proportion of energy is useful for the internal dynamics of
the particle. This translates into a slowing down of time, relative
to absolute time. Other aspects of the theory of relativity can be
explained in classical language as well. However, the main
predictions of special relativity are not supported. This comes from
the fact that the speed of fundamental waves cannot be invariant, as
described. The speed of elementary sources can be locally invariant,
because of the total effect of the waves of other sources on them.
For interpretive purposes, this means that gravity was the reason
for the observed invariance in terrestrial experiments, not the
position of the observer. Moreover, this also means that the speed
of light can not be the same in all contexts, even when measured
from the same inertial frame (though, it matters how the
measurements are made).

According to the preceding comments, the speed of light must be
invariant relative to the predominant source of gravity. This has an
interesting implication that can be verified. In the interplanetary
medium, the Sun should be the predominant source of gravity. Ergo,
the speed of light should be constant in the heliocentric system of
reference, but not in the system of reference of an interplanetary
space probe. If two vessels with synchronized clocks are moving with
the same velocity away from the Sun, one behind the other, their
radio communications should take different amounts of time in each
direction. If the times of emission and detection of each signal are
properly recorded, log comparison should reveal that messages always
arrived faster from the first ship to the second, than from the
second to the first. This is because the second vessel is moving
towards the source of the signal (at the time of its emission),
whereas the first vessel is moving away from it in the inertial
system of the Sun. Considering the number of twin missions to Mars
and Venus, it should not be very difficult to verify this
conjecture. According to several dissident reports on the internet,
such observations have already been made. However, it is preferable
to have some sort of authoritative verification on this matter. If
anyone from the audience has access to relevant information, or is
otherwise qualified to comment on the issue, we would be grateful
for any feed-back that can be afforded.

A different phenomenon that seems to support the same conclusions is
the Pioneer anomaly. This topic has enjoyed a lot of attention among
professionals and in the media. Reported initially by Anderson
\textit{et al.} [5], it concerns the fact that all the space probes
en route to the edges of the Solar system display an unexpected
constant acceleration toward the sun. Recently, this topic has been
in the news again, because the anomaly has been extended to include
several cases of acceleration towards flown-by planets as well [6].
It is important to note that many theories have been proposed to
account for the Pioneer anomaly, but none of them has been accepted
as final yet. Of special interest to our discussion is the analysis
of Teocharis, who has looked at this phenomenon from the point of
view of a model similar to ours. His unpublished paper suggests that
it is possible to overcome the anomaly by assuming that the speed of
light is locally invariant but not universally constant. By taking
into account the possibility that light might travel at different
speeds in different gravitational media, one could remove the
grounds for concluding that the space probes are accelerating.
(References to this and similar unpublished papers can be provided
upon request). This hypothesis has yet to be confirmed, but it was
formulated without knowledge about the model presented above. Its
predictions converge with those of our approach, suggesting
compatibility with the experimental record. Hopefully, the Pioneer
anomaly will be resolved in the near future in a way that will help
us determine the validity of the forgoing considerations.

Another relevant gravitational effect is the inevitable anisotropy
in the field of moving cosmic bodies. In a classical medium, a
moving source must experience a Doppler shift. Hence, the field of
any planet must be asymmetric along several axes, corresponding to
its motion as part of the galaxy, star system, as well as its
orbital motion. Though, it is hard to say if the anisotropy of a
large association of bodies can be detected from inside in a
straightforward manner. The only pattern of motion that is
individual for a planet is its orbital motion. Accordingly, the
gravitational field of all planets must contain a detectable
dawn-dusk asymmetry, which should be stable against all local
sources of variation (such as changes of mass distribution due to
tides, ocean currents, etc.). Such an asymmetry for the planet Earth
should not be difficult to verify, considering the amount of data
that is already available. With the GRACE project [7] successfully
under way, and several other missions in the pipeline, the
hypothesis of a dawn-dusk asymmetry in the field of our planet
should be verified beyond any reasonable doubt. Moreover, the
gravitational field should be stronger on the dawn side (in the
direction of motion), if our model is to be validated. We hope that
scientists with competence in this area of research would respond to
our hypothesis by checking the data for relevant indicators.

It is also important to note that all the planets from the Solar
system should have the mentioned asymmetry in their gravitational
fields. Therefore, the hypothesis is also verifiable with
astronomical observations, by virtue of gravitational lensing. It is
known that Jupiter is massive enough to deflect the apparent
position of distant sources of radiation. Moreover, current
technology allows for very accurate measurements of these effects,
as demonstrated in a recent experiment. Kopeikin and Fomalont [8]
have tried to use their observation for an estimate of the speed of
gravitational waves. Their interpretation was a little ambiguous and
is still open for debate. However, the important fact is that
current technology appears to be able to detect even small effects.
This means that it should be possible to use already existing data,
in order to see if the gravitational lensing of Jupiter varies
before and after eclipsing a source of light. In response to our
questions by e-mail, the authors claimed that their data was not
used to test for the specified type of anisotropy. It would be very
important if someone did perform the necessary analysis, because
Jupiter also has other interesting asymmetries in its fields. For
example, radio images of the gas giant contain remarkable lobes,
which appear anisotropic in most prints that were examined by us.
(See, for example, ref. [9]). This is probably indicative of the
dawn-dusk asymmetry in the magnetic field of Jupiter, which is
independently verified, and which is also consistent with the
implications of our model.

To sum up, the gravitational implications of our take on fundamental
mechanics appear quite odd, compared to other models. However, they
have the virtue of being easily verifiable. In some cases the data
is already available for analysis. At the same time, we do not know
of any experimental fact that contradicts this model. Therefore, it
is a fruitful avenue of research, because it could illuminate many
aspects of fundamental physics without large-scale financial
investments.

\section{Relevant electromagnetic effects}

As suggested in the previous section, moving charged and/or
magnetized bodies should experience deformations of their fields due
to Doppler effects. Moreover, these deformations should be
detectable internally, from their own systems of reference. They
should manifest as field asymmetries. It is not an excessive task to
propose relevant experimental settings, in order to test this
conclusion. However, it is instructive to note that the
magnetosphere of the Earth has a well-studied dawn-dusk asymmetry.
This phenomenon is detectable in the form of diurnal variations in
the geomagnetic declination, as well as via measurements of
azimuthal distributions of the cosmic showers. It is just as
interesting that Jupiter also has a strong dawn-dusk magnetospheric
asymmetry, similar to other planets and moons from the Solar system
that have been studied. According to our interpretation, all cosmic
bodies that have magnetic fields should manifest asymmetries along
their axis of motion. In the case of planets that follow
counterclockwise orbital and axial rotation, the magnetic field
should be stronger on the dawn side. It is important to keep in mind
that magnetospheres are very complex fields, subjected to multiple
causes for variation. Our preliminary surveys of the relevant
literature have not yielded enough information for solid
conclusions. For a proper confirmation, all of those magnetic fields
should be asymmetric in the predicted direction, and they should
also be stable enough to be attributed to constant Doppler effects.
Nevertheless, it is already significant that these asymmetries
exist, and we want to draw more attention to them.

The interpretation of forces in terms of fundamental waves leads to
a new way of understanding electrodynamic phenomena. According to
currently accepted theories, magnetism is not really an independent
force. It is a relativistic effect of moving charged bodies. With
increasing velocity, charge weakens and transforms into magnetism.
This relationship is hard to dispute phenomenologically. However, it
does not seem to work in all cases. For example, it is a known fact
that parallel currents attract, while antiparallel currents repel.
If the electrons from parallel currents move with the same velocity,
they are practically at rest relative to each other. Instead of
magnetic attraction, electric repulsion should be the dominant
effect, but this is not the case. Another important feature of
modern models is the assumption that charge has fundamental
monopoles, and that magnetic monopoles could exist as well. These
implications are in direct contrast with the model presented above.
Under the assumption of fundamental generation, magnetic and
electric waves co-exist at all times, despite the different
macroscopic manifestations. Moreover, those waves can only be
produced in pairs of opposite polarity, propagating in opposite
directions. There can be no magnetic monopoles, or charge monopoles.
The structure of charged subatomic particles is supposed to be such
that electric waves propagate along the direction of motion. Due to
the Doppler effect, the electric field must be denser in the
direction of motion, producing an overall surplus of charge in most
frames of reference. At the same time, the magnetic waves are
orthogonal do the direction of motion and are not distorted like
that. Their bi-polarity is always obvious. This description entails
that static electrons with similar orientation cancel each other's
magnetic effects (in the rest system of reference), whereas their
charge is cumulative. On the other hand, electrons lined
sequentially in currents should cancel most of each other's electric
force, exposing their magnetic force relative to moving targets. In
conclusion, the assumption of fundamental generation implies that
electricity and magnetism do not transform into each other. It is
only their manifestation that depends on relativistic
considerations. In terms of observations, this means that static
configurations of charged particles should have detectable magnetic
effects (as in the example with attracting currents). Another
implication is that currents of charged particles should have
detectable static fields as well. Such effects have been observed in
the past, and they have yet to be conclusively interpreted.

A set of experiments with high voltage discharges, reported by
Podkletnov and Modanese [10], is particularly relevant for this
presentation. On the one hand, the authors showed that large numbers
of electrons, released by a superconducting emitter through a
rarefied gaseous medium, did not produce lightning sparks. They
rather propagated in the form of flat disks, corresponding to the
surface shape of the cathode, all the way towards the anode. The
electric force should have caused the electrons to fly away from
each other, even as they were attracted by the anode, because they
were not in relative motion at emission. On the other hand, the
discharges have also produced some kind of force beams, which
propagated through material obstacles without absorption far beyond
the boundaries of the anode. The beams had measurable effects on
suspended targets, regardless of their electrical properties
(charged and neutral alike). In our opinion, this experiment
confirms the existence of static fields along the direction of
propagation of electrons in currents. It also appears to support the
hypothesis of underlying unity between charge and gravity. Since the
publication of the quoted report, Podkletnov has improved his
experiment, detecting the effects of these static pulses up to a
mile from the site of discharge. He has made public and private
claims to the effect that he was able to measure their speed of
propagation, but that his findings were too odd to be accepted for
publication. The surprise was that the measured speeds were
consistently superluminal, exceeding the speed limit for Einstein
causality by almost two orders of magnitude. It is highly desirable
to have such claims confirmed with independent experiments.
Validation may not require costly developments, as suggested by the
proposal with magnetic pulses, described at the beginning of this
presentation. Until then, we are encouraged by the remarkable
consistency between Podkletnov's findings and the predictions of our
qualitative model.

Another important phenomenon that must be mentioned here is the
Biefeld-Brown effect. It concerns the fact that asymmetric
capacitors display a measurable net force in the direction of the
smaller surface. This tendency can be used to extract useful motion
from static devices. Several years ago, a French enthusiast posted
detailed instructions for several gadgets of this sort, sparking an
internet phenomenon called ``the Lifter Project'' [11]. The term
refers to a simple capacitor, built with a large (yet narrow)
tinfoil cathode and a thin wire anode, stretched around a light
wooden structure. Discharges are prevented by the air gap between
the two components, and the whole device lifts into the air, when
high differences in potential are applied (usually, between 10-30
kV). The cause of levitation, as shown in several experiments posted
on the same site, appears to be the tendency of the charged tinfoil
surface to move towards the wire with opposite polarity (which is
fixed above it on the same frame). The tinfoil lifts the whole
structure up, and even has enough potential left for a small
payload. There are more than 350 registered replications of the
lifter, built by amateurs from various countries. According to some
interpreters, this phenomenon should not be possible, because it
appears to violate the principle of momentum conservation. However,
this appearance is deceiving, in our opinion. The fundamental
particles can never stop their constant motion, while the state of
the medium determines the pattern and direction of motion. In the
absence of physical constraints, subatomic particles will react to
each other's presence until all forces cancel out. Symmetry must be
a final outcome for macroscopic observables, not an inviolable
state. Hence, it cannot apply to capacitors with finite capacitance.
Asymmetric capacitors will necessarily have unequal amounts of
charged entities on each side. With charge being strongest in the
direction of motion, according to our model, all particles should
end up being oriented towards the side with opposite polarity. The
capacitor as a whole is pushed in opposite directions by its two
constituent parts, and the side with more charged particles wins.
Our conclusion is that thrusters and lifters displaying the
Biefeld-Brown effect are similar to boats with two propellers on
opposite sides. The net motion of the boat will be in the direction
of action of the strongest propeller. The principle of energy
conservation is not violated any more than in any other experiment
involving static electricity. If the momentum of subatomic particles
was stored from an external source, as commonly suggested by many
theories, this phenomenon would have been very difficult to explain.
As a corollary, the Biefeld-Brown effect is a strong argument in
favor of the hypothesis of constant and indestructible fundamental
motion.

\section{Relevant quantum effects}

Puzzles and paradoxes enjoy a central role in quantum mechanics.
This is probably why there are so many published reports on various
experiments that study the same phenomenon from different angles.
Accordingly, we had more opportunities to test the implications of
our hypotheses in this area. Our main conclusions on quantum
phenomena have been already presented elsewhere [12], and there is
no time to describe them here in detail. Two important features need
to be mentioned, though. Quantum mechanics is primarily
quantitative, and this may blur important distinctions between
different types of phenomena. For example, many types of
correlations are often treated as related examples of quantum
interference, or entanglement. We find it important to differentiate
between Bell-type correlations and momentum correlations that
produce actual fringes. Bell's inequality is a formal instrument,
which has very specific implications about the type of statistics
that can violate its predictions. The most common interpretation is
that a violation of Bell's inequality rules out realism. Still, if
we assume the existence of a fundamental level of matter with
special rules of interaction, then realist models which violate
classical statistics are still possible. It is sufficient to allow
that entangled quanta violate the Malus law in order to predict
violations of Bell's inequality without non-local interactions. As
mentioned above, our model of fundamental mechanics does violate
Einstein causality by assuming that waves can propagate faster than
light. However, this does not seem to be relevant for polarization
entanglement. The initial state of some pairs of photons can be such
as to produce stronger correlations than any classically polarized
and purified beam. In contrast, interference fringes can only emerge
via physical interactions that occur well after emission. Because of
the assumption that waves must overlap for this type of phenomena,
sources of waves must be close enough for visible effects. Another
difference from alternative interpretations is our insistence on the
fact that coherence is not sufficient for interference. This
consideration enabled us to predict the limits of Young
interference, and to extend our conclusions to the interpretation of
various non-classical phenomena, such as ghost interference, quantum
imaging and quantum erasure. The main advantage of this model is
that classical analogies for quantum interactions are always
possible, and this enables the development of conclusive new tests
for its predictions.

It is always helpful to have experiments that test the indirect
implications of one mechanism or another. However, the holy grail of
experimental physics is to obtain direct verification of any kind of
process. In contrast to earlier assumptions that quantum properties
can not be observed, we would like to propose a simple experiment to
reveal the main attributes of optical interference. As mentioned
repeatedly in this presentation, the photons can be described as
trains of elementary entities, which generate waves. It is the waves
that guide the photons into fringes. Still, it is the sources
(particles) which generate the clicks at the detectors. In other
words, it should be possible to develop an experiment with
overlapping wave-packets, but non-overlapping sources of waves. In
suitable arrangements, classical pulses of light could be detected
with time-resolved quantum detectors, in order to obtain
interference distributions for independent sets of detection events.
What do we mean by that? A coherent laser beam can be chopped into a
pulse with well-defined boundaries. The pulse can then be separated
in two with a 50-50 non-polarizing beam-splitter. The two smaller
pulses can then be suitably guided towards a Young interferometer,
along unequal paths. The beginning of the delayed pulse must fall
very close to the tail of the preceding pulse without overlapping,
such as to make it possible to distinguish one pulse from another in
a time resolved record of detection events. According to our
interpretations, the waves from each pulse could have effects on the
other. This effect should diminish with the square of the distance,
and should be highest within two wave-lengths from any two sources.
Consequently, the photons from the front of the first pulse, or the
tail of the second pulse should not contain any artifacts. However,
the photons that are closest to the boundary between the two pulses
should group into fringes, and the effect should obviously diminish
with distance. Thus, it is possible to test the reality of these
waves and their effects directly. More sophisticated experiments
could even produce a dynamic picture of the process of interference.
It is also important to note that the waves from the second pulse
could only influence the first one if their speed was superluminal.
If these waves are real, but their speed is equal to the speed of
light, only the second pulse should contain fringes. Therefore, the
experiment can test for the reality of waves, and also yield a
general indication about their speed.

\section{Conclusion}

In this presentation we have argued that the Universe must have a
fundamental level, whose properties are determined by
theory-independent processes. This level must contain discrete
elementary particles, producing energy in discrete steps, or else it
leads to philosophical inconsistencies at higher levels of
organization. All the energy in the Unverse must be in the form of
waves, propagating away from the points of excitation, where
particles interact with a fundamental continuous medium. The most
important prediction concerns the speed of those waves. According to
our interpretation, this is the speed of propagation of changes in
magnetic, electric, and gravitational fields. It has to be
superluminal in order to be consistent with known observations, but
the exact value has yet to be determined. The model was also shown
to be consistent with the principle of energy conservation. It does
not contradict the results of the Michelson-Morley experiments.

In our opinion, this approach is very well supported by
interferometric data, as argued in other presentations as well. It
is particularly consistent with several unexplained electromagnetic
phenomena, such as the Biefeld-Brown effect, Podkletnov's
experiments with high voltage discharges, and possibly with the
dawn-dusk asymmetries in the magnetic fields of several planets.
Preliminary reports appear to support the implications for
gravitational effects, even though the evidence is still
inconclusive in this area. Additional research on the Pioneer
anomaly, the hypothesized dawn-dusk asymmetries in the gravity of
the Earth and Jupiter, and other related phenomena, could cover this
gap in the near future. To sum up, we have found a way to simplify
our understanding of fundamental processes in the Universe and to
explain novel recent discoveries that do not always seem to fit
earlier models.

\end{document}